# Spiral cracks without twisting

A fascinating class of patterns, often encountered in nature as wiggling cracks on rocks, dried out fields and tectonic plates, is produced by the fracture of solids[1]. Here we report the observation and modeling of an unusual type of patterns consisting of spiral cracks within fragments of a thin layer of drying precipitate. We find that this symmetry-breaking cracking mode arises naturally not from twisting forces, but from a propagating stress front induced by the fold-up of the fragments.

Many recent studies have analysed the morphology of fractured surfaces and lines[2-6], which typically show a cellular and hierarchical pattern. When twisting forces are applied, spiral-like of fracture is naturally expected. An example is the helical fracture of the tibia suffered in ski accidents[7]. However, spiral cracks are not restricted to those situations, as we illustrate with a very simple experiment of the drying of a fine aqueous suspension of precipitate. As generally expected, during drying the suspension solidifies and later fragments into isolated parts (Fig. 1a). Surprisingly, for very fine precipitates and solidified layer thickness between 0.2 to 0.5 mm, regular spiral as well as circular crack paths show up inside the fragments (Fig. 1b). Depending on the grain size, precipitate type and layer thickness, the size of the spirals vary widely from several hundred microns to a few mm. With naked eye they might look as small dots, but their full beauty is revealed under a microscope (Fig. 1c). The spiral cracks are not limited to one peculiar material, since we managed to produce them in three different precipitates: $Ni_3(PO_4)_2$, $Fe_4[Fe(CN)_6]_3$ and $Fe(OH)_3$.

Careful in-situ observations suggested the following mechanism. The spirals and circle-shaped structures form only after the fragmentation process is over. Due to the humidity gradient across the thickness, the fragments gradually fold up and detach from the substrate, generating large tensile stresses in the radial direction, at and normal to the front of detachment. The extent of the attached area shrinks as the ring-shaped front advances inward due to ongoing desiccation. When the stress at the front exceeds the material strength, a crack is nucleated. Since the nucleation is seldom symmetrical with respect to the boundaries, the crack tends to propagate along the front only in one direction where more stresses can be released. By the time the crack growth completes a cycle, the front has already advanced, leading eventually to an inward spiral crack. Since the stresses are concentrated at the layer-substrate interface, the spiral is confined there, with a typical penetration of 20-60% of the thickness. The patterns being largely spiral suggests that crack propagation is favoured over nucleation, otherwise we would see more cylindrical concentric cracks. Although in a few instances we did observe the latter, the overwhelming majority is the spiral structure.

To test the proposed mechanism, we implement it in a mesoscopic computer model[8], which describes fracture on a frictional substrate. In this model the grains in the layer are represented by blocks on a triangular lattice, interconnected among neighbors by springs. The system is pre-strained, and then relieved quasi-statically in a physical way. The relaxation is dictated by the competition between stick-and-slip and bond breaking. Focusing on the post-fragmentation process, we impose a circular-shaped, inward propagating stress field to mimic the advancing detachment front. In a rather narrow parameter region, the desired spiral cracks are successfully reproduced (Fig. 1d). Tighter-binding spirals can be obtained for smaller penetration depth, in agreement with experiment and the physical expectation based on screening effects in the stress field. Similar evidence of spiral crack patterns under specific conditions has also been reported recently[9].

**K.-t. Leung[*], L. Józsa[†] , M. Ravasz[‡] & Z. Néda[‡§]**

[*]*Institute of Physics, Academia Sinica, Taipei, Taiwan 11529, R.O.C.,*
[†]*Department of Chemistry,* [‡]*Department of Physics, Babes-Bolyai University, Cluj, RO-3400, Romania*
[§]*Department of Physics, University of Notre Dame, Notre Dame, Indiana 46556, USA*

e-mail: *zneda@nd.edu*

**References:**

1. Lawn, B. *Fracture of brittle solids*, 2nd ed. (Cambridge Univ. Press, New York, 1993).
2. Skjeltorp, A.T. & Meakin, P. *Nature* **335**, 424-426 (1988).
3. Yuse, A. & Sano, M. *Nature* **362**, 329-331 (1993).
4. Groisman, A. & Kaplan, E. *Europhys. Lett.* **25**, 415-420 (1994).
5. Bai, T., Pollard D.D. & Gao, H. *Nature* **403**, 753-756 (2000).
6. Shorlin, K.A., de Bruyn, J.R., Graham, M. & Morris, S.W. *Phys. Rev. E* **61**, 6950-6957 (2000).
7. Bostman, O.M. *J. Bone Joint Surg.* (Br) **68**, 462-466 (1986)
8. Leung, K.-t. & Neda, Z. *Phys. Rev. Lett.* **85**, 662-665 (2000)
9. Xia Z.C., Hutchinson J.W. *J. Mech. Phys. Solids* **48**, 1107-1131 (2000)

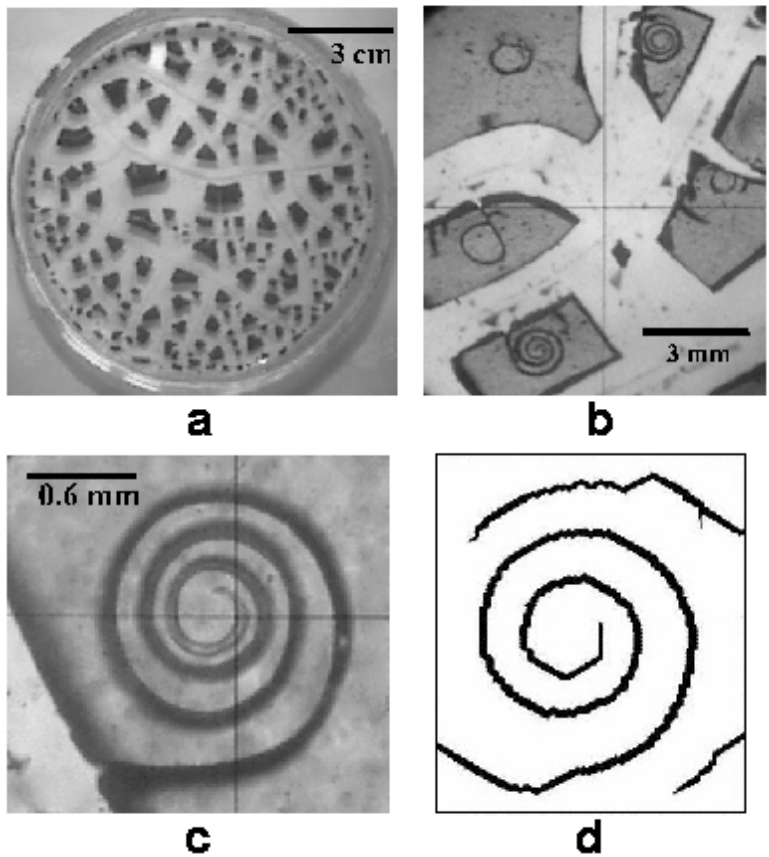

Figure caption:

Figure 1  **a,** Characteristic fragmentation pattern of $Ni_3(PO_4)_2$ precipitates after desiccation in a Petri dish. **b,** Spiral and circular structures obtained inside the fragments. **c,** Close-up of a spiral crack. **d,** Computer simulated spiral crack using a mesoscopic spring-block model.